\documentclass[preprint2]{emulateapj}
\usepackage{graphicx}
\usepackage{url}
\usepackage{bm}

\def\arcsec{\ensuremath{^{\prime\prime}}}

\def\Av{\mbox{$A_V$}}
\def\Rv{\mbox{$R_V$}}

\newcommand{\Wilson}{UDS10Wil}
\newcommand{\CCSN}{CC\,SN}

\newcommand{\SNIa}{SN\,Ia}

\newcommand{\JHU}{Department of Physics and Astronomy, The Johns Hopkins University, Baltimore, MD 21218.}
\newcommand{\STScI}{Space Telescope Science Institute, Baltimore, MD 21218.}
\newcommand{\Berkeley}{Department of Astronomy, University of California, Berkeley, CA 94720-3411.}
\newcommand{\Riverside}{Department of Physics and Astronomy, University of California, Riverside, CA 92521.}
\newcommand{\WKU}{Department of Physics, Western Kentucky University, Bowling Green, KY 42101.}
\newcommand{\Copenhagen}{Dark Cosmology Centre, Niels Bohr Institute, University of Copenhagen, Juliane Maries Vej 30, DK-2100 Copenhagen, Denmark.}
\newcommand{\Arizona}{Department of Astronomy, University of Arizona, Tucson, AZ 85721.}

\newcommand{\NotreDame}{Department of Physics, University of Notre Dame, Notre Dame, IN 46556.}
\newcommand{\TelAviv}{School of Physics and Astronomy, Tel-Aviv University, Tel-Aviv 69978, Israel.}
\newcommand{\Rutgers}{Department of Physics and Astronomy, Rutgers, The State University of New Jersey, Piscataway, NJ 08854.}
\newcommand{\CfA}{Harvard/Smithsonian Center for Astrophysics, Cambridge, MA 02138.}

\newcommand{\ALMA}{Joint ALMA Observatory, ESO, Santiago, Chile.}
\newcommand{\AMNH}{Department of Astrophysics, American Museum of Natural History, Central Park West and 79th Street, New York, NY 10024-5192.}

\ifx\pdfoutput\undefined
  \pdffalse
  \DeclareGraphicsExtensions{.eps,.ps}
\else
  \ifnum\pdfoutput=1
    \DeclareGraphicsExtensions{.pdf,.png,.jpg}
  \else
    \DeclareGraphicsExtensions{.eps,.ps}
  \fi
\fi

\begin{document}

\title{The Discovery of the Most Distant Known Type Ia Supernova at Redshift
  1.914}

\author{David O. Jones\altaffilmark{1}, Steven A. Rodney\altaffilmark{1,2},
  Adam G. Riess\altaffilmark{1,3}, Bahram Mobasher\altaffilmark{4}, Tomas
  Dahlen\altaffilmark{3}, Curtis McCully\altaffilmark{5}, Teddy F. Frederiksen\altaffilmark{6}, Stefano
Casertano\altaffilmark{3}, Jens
Hjorth\altaffilmark{6}, Charles R. Keeton\altaffilmark{5}, Anton
Koekemoer\altaffilmark{3}, Louis-Gregory
  Strolger\altaffilmark{7}, Tommy G. Wiklind\altaffilmark{8}, Peter Challis\altaffilmark{9},
Or Graur\altaffilmark{10,11}, Brian Hayden\altaffilmark{12}, Brandon
Patel\altaffilmark{5}, Benjamin J. Weiner\altaffilmark{13},
Alexei V. Filippenko\altaffilmark{14}, Peter Garnavich\altaffilmark{12}, 
Saurabh W. Jha\altaffilmark{5}, Robert P. Kirshner\altaffilmark{9}, Henry C.
Ferguson\altaffilmark{3}, Norman A. Grogin\altaffilmark{3}, and Dale Kocevski\altaffilmark{9}}

\altaffiltext{1}{\JHU}
\altaffiltext{2}{Hubble Fellow.}
\altaffiltext{3}{\STScI}
\altaffiltext{4}{\Riverside}
\altaffiltext{5}{\Rutgers}
\altaffiltext{6}{\Copenhagen}
\altaffiltext{7}{\WKU}
\altaffiltext{8}{\ALMA}
\altaffiltext{9}{\CfA}
\altaffiltext{10}{\TelAviv}
\altaffiltext{11}{\AMNH}
\altaffiltext{12}{\NotreDame}
\altaffiltext{13}{\Arizona}
\altaffiltext{14}{\Berkeley}

\begin{abstract}

We present the discovery of a Type Ia supernova (SN) at redshift $z = 
1.914$ from the CANDELS multi-cycle treasury program on the \textit{Hubble
  Space Telescope (HST)}.  This SN was discovered in the infrared
using the Wide-Field Camera 3, and it is the highest-redshift Type Ia SN
yet observed.  We classify this object as a SN\,Ia by comparing its light
curve and spectrum with those of a large sample of Type Ia and
core-collapse SNe.  Its apparent magnitude is consistent
with that expected from the $\Lambda$CDM concordance cosmology.  We 
discuss the use of spectral
evidence for classification of $z > 1.5$ SNe\,Ia using {\it HST} grism
simulations, finding that spectral data alone can
frequently rule out SNe\,II, but distinguishing between SNe\,Ia and
SNe\,Ib/c can require prohibitively long exposures.  In such
cases, a quantitative analysis of the light curve may be necessary for
classification.  Our photometric and spectroscopic classification methods 
can aid the determination of SN rates and cosmological parameters from the full
high-redshift CANDELS SN sample.

\end{abstract}

\section{Introduction}

Over the past decade, measurements of Type Ia supernovae (SNe) at redshift
$z \gtrsim 1$ have extended the observed population to a time when the universe
was matter dominated \citep{Riess01,Riess04,Riess07,Suzuki12,Rodney12,Rubin12}.  At 
these lookback times of $\gtrsim 7$\,Gyr, the predicted effects of dark energy
are small, while the typical conditions under which SNe form 
are increasingly different from local environments.  

These characteristics may allow observations at high redshift to
constrain an evolutionary change in SN\,Ia brightness independent of our
understanding of dark energy.  This type of systematic shift in
magnitude could be caused by
changing metallicity or progenitor masses (e.g., \citealp{Dominguez01}).  
Such an effect could be present at a lower level in
intermediate-redshift SN samples  ($0.2 \lesssim z \lesssim 1.0$), 
and therefore be a 
source of uncertainty in the determination of the dark energy 
equation-of-state parameter ($w=P/(\rho c^2$); \citealp{Riess06}).

Observations of high-redshift SNe\,Ia could also place constraints on
the binary companions of SN progenitors.  The two most likely SN\,Ia
progenitor models are the
single-degenerate scenario, where a white dwarf accretes matter from a
main-sequence or giant companion, and the double-degenerate scenario, where
SNe occur through the merging of two carbon-oxygen (C-O) white dwarfs.  A 
substantial difference between these mechanisms, however, is
the typical time interval from progenitor formation to explosion; 
progenitors would likely take
$\gtrsim 10^9$\,yr to reach the Chandrasekhar limit by mass transfer
from a nondegenerate companion, but would more often take less time 
in a system of two C-O white dwarfs
(for a recent review of SN\,Ia progenitors, see \citealp{Wang12}).  The
distribution of times between formation and explosion, known as the delay-time
distribution (DTD), can therefore be used to set constraints on SN
progenitor models.  Observations of SN rates measure the
convolution of the DTD with the cosmic star-formation history,
and high-redshift rates are the most sensitive to delay times \citep{Strolger10,Graur11}.

Due to the high sensitivity and angular resolution of the 
\textit{Hubble Space Telescope (HST)}, its 
Advanced Camera for Surveys (ACS) has been the most effective
instrument for observing and monitoring SNe out to $z \approx 1.5$.  
To find SNe at higher redshifts in the rest-frame optical, where they 
are brightest and we understand them best, searching in the near-infrared 
(IR) with the recently installed Wide-Field Camera 3 (WFC3) allows
SN surveys to reach unprecedented depths not accessible from
the ground (F160W limiting Vega magnitude $\sim 25.5$, equal to
the peak observed brightness of a typical SN\,Ia at $z \approx 2.5$).

\begin{figure}
\includegraphics[width=85mm, height=85mm]{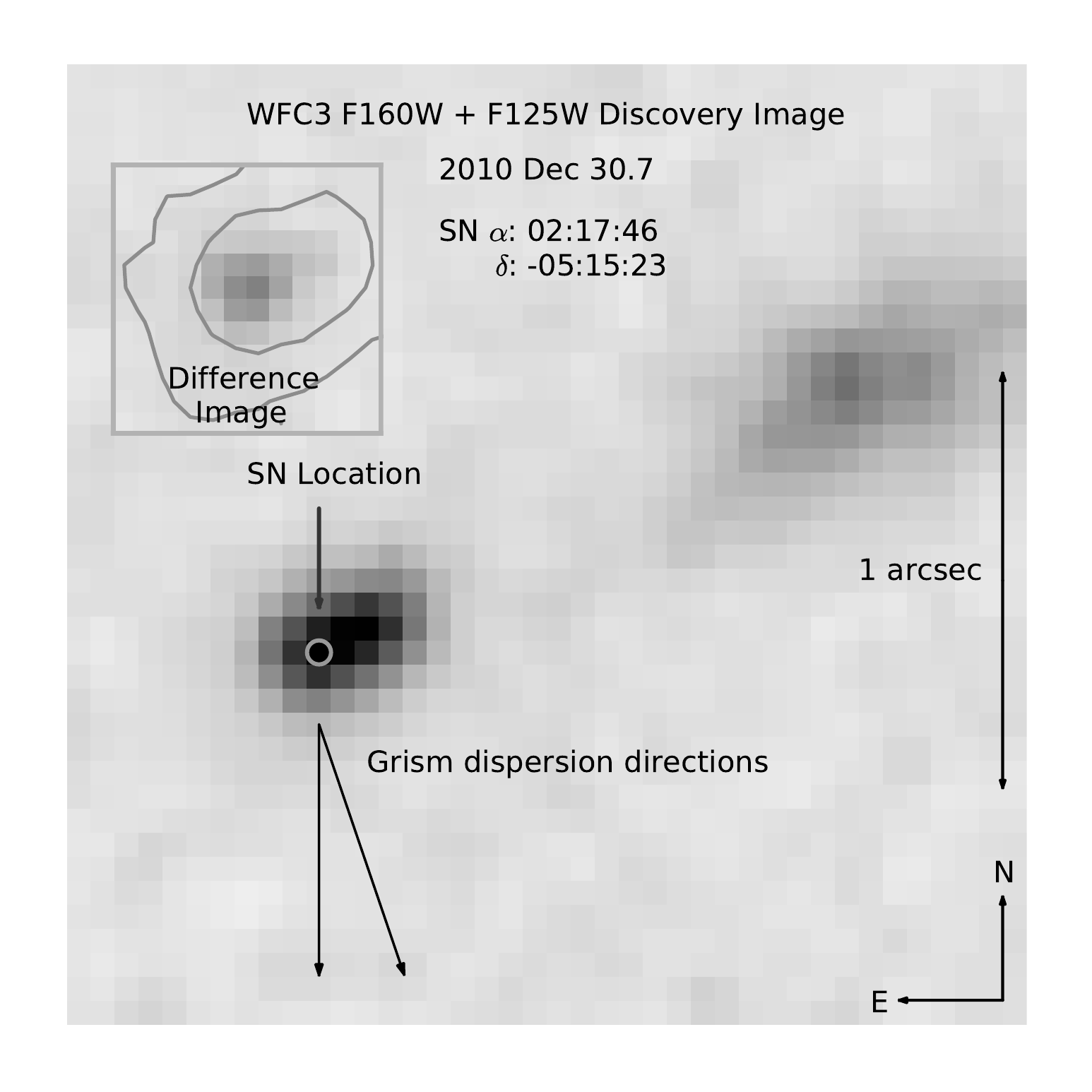}
\caption{The WFC3 F160W + F125W discovery and difference images (using a
  late-time, SN-free template) for SN \Wilson.  The SN is located
  $\sim 0$\arcsec$.1$ from the center of the host galaxy (2 ACS
  pixels).  The contours plotted on the difference image of the SN (upper left)
  show the regions containing 68\% and 95\% of the host galaxy light.
  The center of the nearest neighboring galaxy, which causes minimal 
  lensing of the SN (see \S4.1),
  is located $\sim 1$\arcsec$.5$ away.}
\label{fig:discovery}
\end{figure}

In this paper we present observations of a SN\,Ia at $z=1.91$ (SN \Wilson),
the highest-redshift SN\,Ia discovered to date.  It was found in
the Cosmic Assembly
Near-infrared Deep Extragalactic Legacy Survey (CANDELS, PI: Faber \&
Ferguson; \citealp{Grogin11,Koekemoer11}).  The full CANDELS SN sample is 
designed to measure SN rates and to study SN systematics at redshifts greater than 1.5.
Similar to SN Primo, a $z = 1.55$ WFC3-discovered SN \citep{Rodney12,Frederiksen12}, \Wilson\ also
has spectroscopic evidence for classification.  We present the discovery of SN
\Wilson\ in \S2.
Section 3 discusses its classification from photometry and {\it HST}
 grism spectroscopy.  In \S4 we estimate the brightness 
correction due to gravitational lensing and fit the light curve.
We discuss our results and the {\it HST} spectral classification
in \S5, and our conclusions are given in \S6.

\section{Discovery}

\begin{deluxetable*}{lllrrc}
\tablewidth{0pt}
\tablewidth{0pt}
\tablecolumns{10}
\tabletypesize{\scriptsize}
\tablecaption{Photometric Observations}
\renewcommand{\arraystretch}{1}
\tablehead{
\colhead{UT Date}&
\colhead{MJD}&
\colhead{Filter}&
\colhead{Exposure Time}&
\colhead{Flux (counts s$^{-1}$)}&
\colhead{Vega Mag}}
\startdata
2010 Nov. 08.8 & 55508.1 & F814W & 3517.0 & 0.143   $\pm$   0.054  &    27.635 $\pm$  0.413\\
2010 Nov. 11.2 & 55511.2 & F160W & 1605.8 & 0.517  $\pm$    0.074  &    25.221 $\pm$  0.156\\
2010 Nov. 11.2 & 55511.2 & F125W & 955.9 & 0.698  $\pm$    0.096  &    25.535 $\pm$ 0.149\\
2010 Dec. 28.0\tablenotemark{*} & 55557.4 & F814W & 3817.0 & $-$0.063  $\pm$    0.041 & ...    \\
2010 Dec. 30.7\tablenotemark{*} & 55560.7 & F160W & 1705.9 & 1.22  $\pm$    0.079 &     24.290 $\pm$  0.070\\
2010 Dec. 30.8\tablenotemark{*} & 55560.8 & F125W & 955.9 & 1.403  $\pm$    0.102 &     24.776 $\pm$  0.079\\
2011 Jan. 12.6 & 55573.6 & F160W & 3617.6 & 0.901  $\pm$    0.063 &     24.616 $\pm$  0.076\\
2011 Jan. 12.8 & 55573.8 & F125W & 3617.6 & 0.759  $\pm$    0.062  &    25.443 $\pm$  0.089\\
2011 Jan. 13.6 & 55574.6 & F850LP& 1994.0 &  $-$0.018  $\pm$    0.035 & ...    \\
2011 Jan. 23.4 & 55584.3 & F160W & 3667.6 & 0.780  $\pm$    0.061 &     24.774 $\pm$  0.085\\
2011 Jan. 23.4 & 55584.4 & F125W & 3867.6 & 0.535  $\pm$    0.059  &    25.823 $\pm$  0.118\\
2011 Feb. 04.2 & 55596.1 & F160W & 3767.6 & 0.441  $\pm$    0.061 &     25.392 $\pm$  0.150\\
2011 Feb. 04.2 & 55596.1 & F125W & 3717.6 & 0.437  $\pm$    0.062 &     26.043 $\pm$  0.154\\
2011 Feb. 16.1 & 55608.1 & F160W & 4973.5 & 0.309  $\pm$    0.058 &     25.779 $\pm$  0.205\\
2011 Feb. 16.3 & 55608.2 & F125W & 4973.5 & 0.183  $\pm$    0.057 &     26.989 $\pm$ 0.337\\
\\
2011 Jan. 12.7 & 55573.7 & G141 & 39088.0 & (grism obs) & ...\\
\enddata
\tablenotetext{*}{Discovery epoch.}
\label{table:photometry}
\end{deluxetable*}

SN \Wilson\ was discovered in the second epoch of CANDELS observations of 
the UKIDSS Ultra-Deep Survey field (UDS;
\citealp{Lawrence07,Cirasuolo07}) on 2010 December 30, after subtracting
the images obtained in the first epoch (2010 November 11).
It was detected at high significance in both F160W and F125W
difference images, while a flux decrement was seen at the same location
in the ACS filter F814W difference image (detected at $\sim 2.5\sigma$).  
The SN searching is 
performed by eye in the difference images, and in this case we could
only subtract the first epoch of UDS observations (50 days before) from
the second epoch, as no earlier WFC3 data were available.  The F814W flux decrement 
suggests that pre-maximum SN light was present in the first
epoch of UDS observations.  Thus, the SN was brighter in the
pre-maximum, shorter-wavelength ACS imaging.  

The WFC3 (F125W $+$ F160W) discovery-epoch image of SN \Wilson\ is shown
in Figure \ref{fig:discovery}, using a late-time (2011 December), SN-free template for
the difference imaging.  The J2000 SN coordinates are
$\alpha = 02^{\rm h}17^{\rm m}46^{\rm s}$, $\delta = -05^\circ 15' 23''$.
It was $\sim 0$\arcsec$.1$\ from the center of its host galaxy 
($\sim 2$ ACS pixels, $\sim 0.9$ kpc in distance), making it highly 
probable that this galaxy was the host and unlikely that the
object was an active galactic nucleus.

At the time of discovery, we determined the photometric redshift of the host 
galaxy to be $> 1.5$, although this was measured before SN-free WFC3
host-galaxy images were available.  At this redshift, the SN colors
(F160W $-$ F814W 3$\sigma$ upper limit, and F125W $-$ F160W) were
consistent with those expected for a SN\,Ia at $z > 1.5$ and inconsistent with a
core-collapse (CC) SN, so we triggered follow-up
observations with the X-shooter spectrograph 
on the ESO Very Large Telescope (VLT) to 
obtain a spectroscopic redshift of the host.\footnote{Based on observations made with ESO telescopes at the La Silla Paranal Observatory under program IDs 086.A-0660 and 089.A-0739.}

  Moreover, we
monitored the SN with the {\it HST} SN Multi-Cycle Treasury follow-up program
(GO-12099; PI: Riess).  We imaged the SN with {\it HST} (20 orbits, to
obtain the light curve as well as SN-free template observations) and
we obtained G141 grism spectroscopy (15 orbits, for resolution $R \approx 
130$).  

To measure the photometry of the SN, we subtracted the late-time
template images from the UDS/SN follow-up observations.  We measured the flux within a fixed aperture of 3-pixel
radius and estimated errors in the flux from the sky noise of the nearby
background-subtracted image.  Details of the {\it HST} observations 
are listed in Table \ref{table:photometry}, and the grism spectrum
is discussed along with the SN classification in \S\ref{sec:grism}.

\subsection{Redshift}

We remeasured the spectral energy distribution (SED) of the SN \Wilson\ host 
galaxy, including photometry from late-time WFC3 and ACS templates as well 
as Subaru, UKIRT, and IRAC data.  The Balmer break is
between the Subaru $z$ band and the WFC3 $J$ band, making the most likely 
redshift between 1.8 and 2.2 (see the lower-left panel of Fig. \ref{fig:spec}).  
Using the X-shooter spectrum, we
narrowed this result by identifying [\ion{O}{2}] and [\ion{O}{3}] doublets 
in the host-galaxy spectrum, giving 
a precise redshift of 1.914.

The result is also consistent with the {\it HST} G141 grism
spectrum, which contains a clear detection of [\ion{O}{3}] 
$\lambda\lambda$4959, 5007. However, the grism spectrum cannot resolve the doublet, 
as the spectrum is convolved with both the shape of the host galaxy and the 
point-spread function (PSF; combined full width at half-maximum intensity 
$\sim 116$\,\AA) and sampled at a resolution of only 46.5\,\AA\,pixel$^{-1}$.  
The VLT spectrum, along with an analysis of the late-type host
galaxy of SN \Wilson, will be presented by Frederiksen et al. (in preparation).

\section{Classification}

We classified SN \Wilson\ by analyzing its light curve and spectrum,
informed by the host redshift.  
As detailed below, we first examined the light curve, finding that it is 
consistent only with a SN\,Ia.  In particular, the combination of its
early-time colors with its rapid late-time decline rate does not agree with
CC SN models.  We then used the spectrum to independently rule
out SNe\,II.  While the spectral absorption features alone are unable to 
convincingly distinguish between a SN\,Ia and a SN\,Ib/c,
SNe\,II have features that are inconsistent with the data (see \citealp{Filippenko97} for a review of SN spectra).

\subsection{Photometric Classification}

To classify SN \Wilson\ we compared the
observed \Wilson\ light curve against Monte Carlo simulations of
Type Ia and CC SNe at redshift 1.91, generated with the SuperNova ANAlysis
software (SNANA\footnote{\url{http://sdssdp62.fnal.gov/sdsssn/SNANA-PUBLIC/ 
.}}; \citealp{Kessler09}).  We used a least-squares fit to scale the magnitude of the simulated
light curves to match our data, thus allowing us to examine how our
data compare to the shapes and colors of simulated SNe while removing any assumptions on cosmology or intrinsic SN
luminosity.  We then measured the $\chi^2$ statistic for each simulated SN compared to
our data and converted these $\chi^2$ values into a Type Ia SN
classification probability using a simple Bayesian framework, similar
to \citet{Poznanski07}, \citet{Kuznetsova07}, and \citet{Sako11}.  The
full Bayesian formalism, along with a description of the simulations
and our Bayesian priors, is presented in the Appendix.  

\begin{figure*}
\plotone{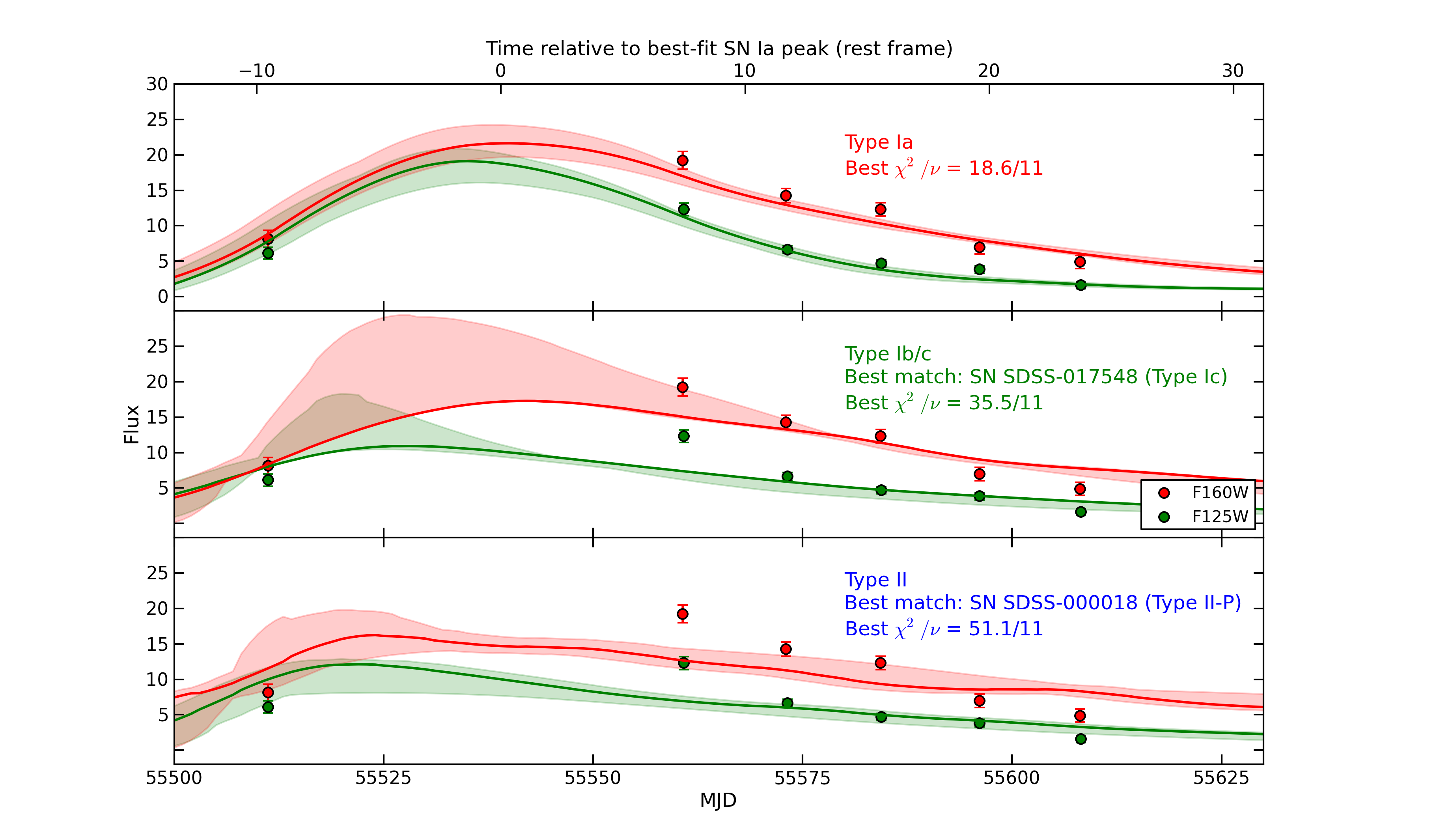}
\caption{The best-fit light curves using SNANA-simulated Type Ia,
Type Ib/c, and Type II SNe (top to bottom).  Red and green colors 
indicate photometry and simulations for the F160W and F125W filters
(approximately rest-frame $B$ and $V$), 
respectively.  Shaded regions
include the range of SN fits that encompass 95\% of the Bayesian
evidence for a given SN type.  Type Ib/c and Type II fits have higher 
$\chi^2$ values (even including model uncertainties), due to their inability 
to match the combination of the colors nearest to maximum light and the rapid rate 
of decline.  The sum
of the evidence for a SN\,Ia model gives a nearly 100\% probability that the
SN is of Type Ia.  We included F814W and F850LP data
in the $\chi^2$ fitting, but have omitted them in this figure for
visual clarity.  The fluxes shown are
transformed from Vega magnitudes using a zeropoint of 27.5.}
\label{fig:lc}
\end{figure*}

Our procedure gives us a very high probability that SN \Wilson\ is a
SN\,Ia.  As such, varying our priors on parameters such as shape,
color, $A_V$, $R_V$, or SN rates has a very minor effect on the outcome.
The reliance on only 43 \CCSN\ templates constitutes the
greatest uncertainty in our procedure.  However,
using a classification procedure very similar to ours, \citet{Sako11} 
found that classification using only 8 CC\,SN templates still returns SN\,Ia classification
purities of $\gtrsim 90$\%.  

We find that the probability of a SN\,Ia was 99.98\%, with
a SN\,Ib/c probability of $2.1\times10^{-4}$ (ruled out at $\sim$3.7$\sigma$) and
a SN\,II probability of $1.0\times10^{-7}$ (ruled out at $\sim$5.3$\sigma$).  This indicates 
that the Type Ia model
dominates the probability calculation, and no \CCSN\ models can
adequately describe the observed photometric data.

Figure \ref{fig:lc} shows the best-fit light curves, along with the
flux range of simulated SN light curves encompassing 95\% of the
Bayesian evidence for each SN type.  The best-fit light curves for
Types Ia, Ib/c, and II SNe return $\chi^2/\nu$ values of 18.6/11,
35.5/11, and 51.1/11, respectively.  Note that these $\chi^2$ values are 
only illustrative of
the quality of the match for each model. They represent the best
match from a large but limited number of random simulations, so one
cannot use these values in $\chi^2$ goodness-of-fit tests for model
rejection. By contrast, the final classification probability relies on the weight of evidence from all realizations of
each model.

Our best-fit $x_1$ and $C$ values for the Type Ia model were 
$-1.56$ and $-0.12$, respectively.  These values are fully
consistent with the SALT2 parameters derived from light-curve
fitting in \S\ref{sec:lightcurve} ($x_1=-1.50\pm0.51$ and
$C=-0.07\pm0.11$).  We note that if we increase the
errors such that the SN\,Ia $\chi^2/\nu \approx 1$ (accounting for the possibility
that we underestimated the uncertainties), the Type Ia probability is still
as high as 99.84\%.  Figure \ref{fig:lc} shows that the nearly 100\% probability of
classification as a SN\,Ia (and the superior best-fit $\chi^2$ value)
arises because the SN\,Ib/c and SN\,II light-curve fits are unable
to match the combination of SN \Wilson's high signal-to-noise ratio (S/N) discovery-epoch colors and its
rapid light-curve decline rate.  

As a verification of this light-curve classification, we used
the Photometric SuperNova IDentification software
(PSNID; \citealp{Sako08}), finding that it also prefers a
SN\,Ia with a slightly higher 4.1$\sigma$ confidence.  The difference
in probability is primarily due to our conservative CC model
uncertainties (see the Appendix), which reduce the $\chi^2$ values of
CC\,SNe.  Although it only uses
8 CC\,SNe, the purity of PSNID classifications has been robustly
tested using Sloan Digital Sky Survey (SDSS) SNe, and it obtained the highest figure of merit in
the SN Photometric Classification Challenge \citep{Kessler10}.

\label{sec:colors}

\subsection{Spectrum}
\label{sec:grism}

\begin{figure*}
\plotone{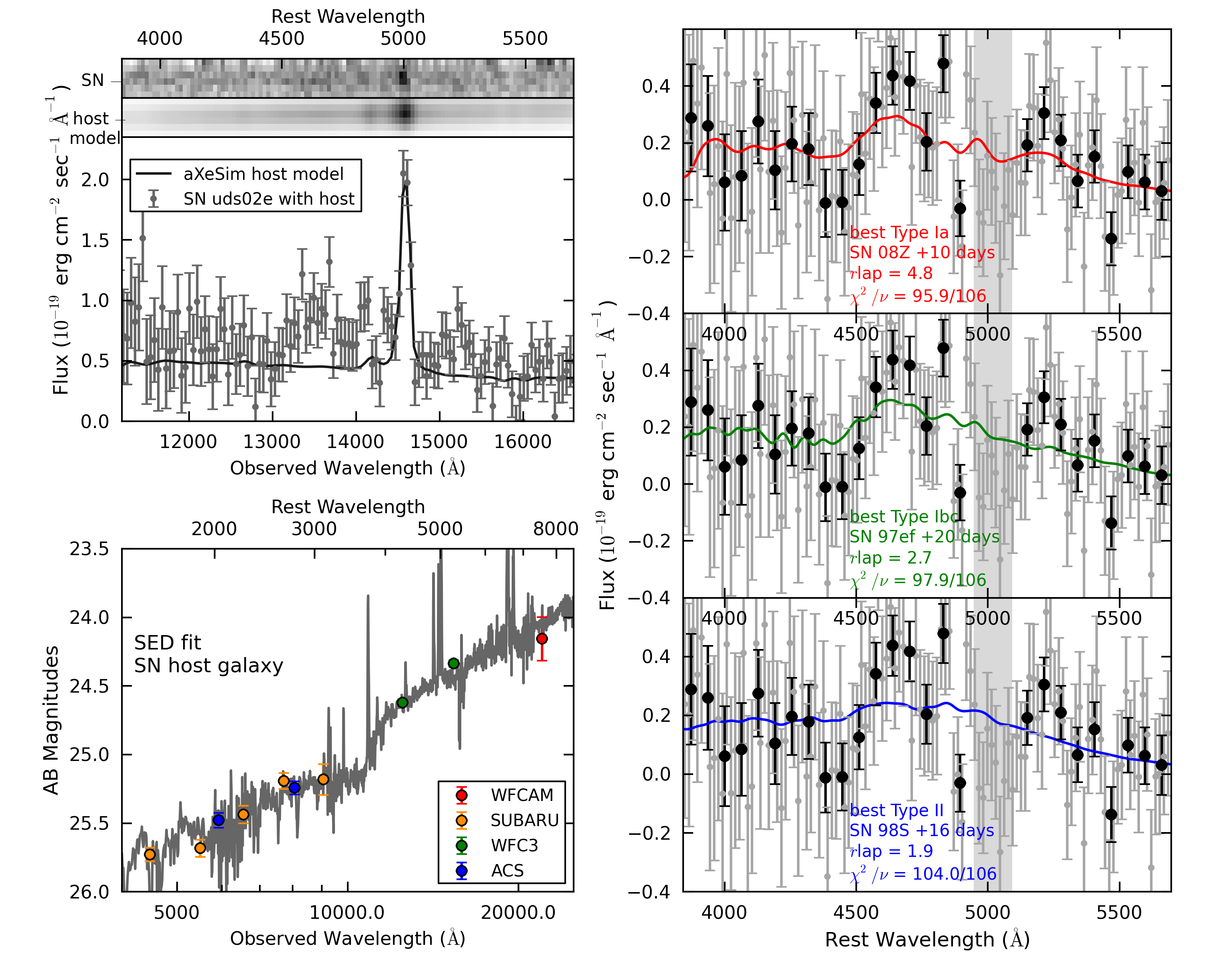}
\caption{The spectrum of SN \Wilson, from which the host-galaxy light
  has been subtracted using a spectrum fit to the host's SED.  The
  host SED and its best-fit spectral template is shown at the bottom left.  
  The two-dimensional and one-dimensional 
  grism spectrum along with the simulated host-galaxy 
  spectrum (the SED convolved with the shape of the host using
  aXeSim) are given at the upper left.  On the right we show the best-fit
  SN templates from SNID for Types Ia, Ib/c, and II.  We illustrate median
  bins to visually emphasize the spectral features, but have used
  the unbinned data for our analysis.  Although
  the spectrum can be fit reasonably well by both a SN\,Ia or a
  SN\,Ib/c, a SN\,II does not contain the spectral
  features seen at $\sim 4600$ and $\sim 5200$\,\AA\ in the rest frame.
  Data in the [\ion{O}{3}] subtraction region (the shaded region)
  were not used in the fit.}
\label{fig:spec}
\end{figure*}

Spectroscopic confirmation of SNe has proven
challenging at these redshifts (see the discussions in \citealp{Rodney12}
and \citealp{Rubin12}), due to the difficulty of obtaining high
S/N observations and the paucity of defining features in the
available window (for \Wilson, $\sim 1.12$--1.65\,$\mu$m; rest frame 
$\sim 3840$--5660\,\AA).  In the case of SN \Wilson, the SN was
separated from its host galaxy by only $\sim 0$\arcsec$.1$,
contaminating the SN spectrum
with host-galaxy light.  We removed the host galaxy from the spectrum by subtracting a section of
the galaxy free from SN light, but the
combined noise from the SN and host-galaxy spectra made a spectral
classification inconclusive, even
with substantial host-galaxy smoothing.

As an alternative approach that avoids adding additional host-galaxy noise to the SN spectrum, we
generated a noise-free synthetic host spectrum.  We fit SEDs, 
using a library of spectral templates, to optical and near-IR Subaru, ACS, 
WFC3, and UKIRT host-galaxy photometry following \citet{Dahlen10}.  We 
then simulated the observed grism host spectrum with the aXeSim 
software package.\footnote{\url{http://axe.stsci.edu/axesim/ .}}
The aXeSim software convolves
the SED with the shape of the host galaxy and {\it HST} PSF and samples the spectrum at
the G141 spectral resolution of 46.5\,\AA\,pixel$^{-1}$.

One would not necessarily expect emission lines to be the same strength 
in the template as in the real galaxy due to its differing metallicity, 
star formation rate, and population of massive stars.  Therefore,
we replaced the pixels covering the [\ion{O}{3}] line in our 
simulated host galaxy with those covering the prominent [\ion{O}{3}] line
from the grism spectrum.  We omitted these pixels (the shaded region in
Fig. \ref{fig:spec}) when we later fit spectral templates to the SN
spectrum, as we did not have a SN-free line measurement to
subtract from the observations.  We then rescaled the
aXeSim output spectrum to match the F160W magnitude of the host galaxy
as measured in the last epoch of follow-up imaging after the SN had
faded.  Our simulated host-galaxy spectrum is shown
in Figure \ref{fig:spec} (upper left).

After subtracting the host-galaxy model from the SN spectrum
contaminated with host light, we used the Supernova Identification
(SNID) code\footnote{\url{http://www.oamp.fr/people/blondin/software/snid/index.html .}} 
from \citet{Blondin07} to match the \Wilson\ spectrum with
Type Ia, Type Ib/c, and Type II SN template spectra to determine 
the best-fit spectrum for each class.  For SN\,Ia fits, we only used
templates within $\pm 3$ rest-frame days of the age of the 
SN \Wilson\ spectrum.  The age is based on the
SALT2 fit in \S4.2, which gives $\sim 12 \pm 1$ day after maximum (rest frame).  For
CC\,SN fits, we used any templates which put the time of maximum between
the first two epochs of observation (the same as our prior in \S3.1).  
When fitting the
spectrum to templates, SNID removes the continuum using a high-order
polynomial fit and only matches the spectral features, making the
fit independent of reddening and brightness.

SNID returns the \textit{r}lap
parameter, which is meant to quantify the quality of the correlation
(see \citealp{Blondin07} for details).  \citet{Blondin07} suggest that 
\textit{r}lap values less than 5 are
inconclusive.  Note that SNID does not apply a
broadening symmetric function \citep{Tonry79}, which should be used when the widths of SN
features are comparable to the resolution of the spectrum.  As the
rest-frame G141 resolution is $\sim 16$\,\AA\ pixel$^{-1}$ and
significant SN features have a typical width $\sim 50$\,\AA, this is not
a major concern. However, the inclusion of this function for grism data
could improve future SNID classifications, especially those at lower redshift.

The right side of Figure \ref{fig:spec} shows the best-fit Type Ia,
Ib/c, and II SN templates with \textit{r}lap values of 4.8, 2.7, and 1.9,
respectively.  We show median bins to emphasize the spectral features,
but we fit spectra to the unbinned data.  The data can be fit by 6 other
normal SNe\,Ia with $r$lap values of at least
4.  Five other SN\,Ib/c fits have an $r$lap of at least 2, and
only two other SN\,II fits have an $r$lap of greater than 1.5.

The $\chi^2$ values for the fits, now including continuum, to Type Ia, 
Ib/c, and II SNe are 95.9, 97.9, and 104.0 (respectively) with 106 
degrees of freedom.  We note that both SNe\,Ia and SNe\,Ib/c can provide
good fits, although the former give a slightly better match.  However,
all SN\,II templates yield a poor correlation; the rest-frame features at $\sim 4600$
and $\sim 5200$\,\AA\ (which are created by neighboring \ion{Fe}{2},
\ion{Fe}{3}, \ion{Si}{2}, and \ion{S}{2} absorption in SNe\,Ia; \citealp{Filippenko97}) 
are not well fit by the spectra of SNe\,II.  We note that although not
all of the SNID SNe\,II are as featureless as the best-fit spectrum
shown in Figure \ref{fig:spec}, they all have difficulty
matching the strength or location of the spectral features.  Even Type
II-P templates, which typically have stronger features, have a maximum
$r$lap of only 1.4.

Because the best $r$lap value is less than 5, this SN cannot be
considered to be spectroscopically confirmed.  In addition, the
``lap'' parameter, which describes the overlap in wavelength space
between the template and SN spectrum, is 0.39 for each of the
best-fit SN\,Ia and CC\,SN templates.  This is below the minimum lap 
of 0.4 used by \citet{Blondin07} for
spectral confirmation.  However, the SN \Wilson\ spectrum still
favors classification as a SN\,Ia and its $r$lap is comparable to
that of other high-redshift SNe\,Ia.  SN Primo
(\citealp{Rodney12}; $z = 1.55$)
had an $r$lap of only 3.7.  We also fit the spectrum of SN SCP-0401
(\citealp{Rubin12}; $z = 1.71$), finding that it is
best matched by featureless SNID Type Ia and Type Ib/c spectra.
However, if we require a match to at least one spectral feature or a
lap value greater than 0.1, the
maximum SN\,Ia $r$lap is 4.6 (SN
1993ac, $+7$ days; lap of 0.2).  SNID
templates begin at $\lambda_{\rm rest} = 2500$\,\AA, so
the first $\sim 500$\,\AA\ of the SN SCP-0401 spectrum were not included
in the SNID fit.

\section{Analysis}

Taken together, the photometric evidence suggests
that \Wilson\ is a SN\,Ia with high confidence.  The spectroscopic
evidence independently favors a Type Ia classification.  We now proceed to
derive its shape and color-corrected magnitude, taking into account the possibility that 
the SN light has been gravitationally lensed by foreground structure.

\subsection{Lensing}

Our ability to use SNe\,Ia as accurate distance indicators to 
constrain cosmological parameters requires us to determine the impact
of foreground-matter inhomogeneities on the flux of the SN (e.g., 
\citealp{Jonsson06}). Even 
without multiple images, gravitational lensing can significantly 
magnify the SN, altering our measurement of its distance.  
SN \Wilson\ is close 
in projection to another galaxy (see Fig. \ref{fig:discovery}) separated 
by only 1\arcsec$.54$.  Therefore, it is necessary to estimate the 
possible magnification of the SN which could lead to a bias in the
derived SN\,Ia distance.  All other foreground sources are greater than
4\arcsec$.5$ away and cause negligible magnification.

We fit the SED of the lens galaxy as 
described by \citet{Wiklind08} to characterize its physical 
properties.  We used a Chabrier initial mass function (IMF; \citealp{Chabrier03})
rather than the Salpeter IMF cited
by \citet{Wiklind08}; the former gives a slightly smaller stellar mass but
is a more accepted model.  To account for 
photometric uncertainties, we drew Monte Carlo samples for 
the measured photometry of the galaxy and used the best-fitting 
SED models to characterize the SED. The SED fit indicates a 
low-mass galaxy with a photometric redshift $0.283 \pm 0.080$ 
and a stellar mass $\log(M_\star/M_{\sun}) = 7.968 \pm 0.222$.   
We used these parameters to create a plausible mass model of 
the galaxy and estimate the magnification of the SN.  

We modeled the
stellar component of the galaxy as an exponential 
disk using parameters measured from GALFIT \citep{Peng02} 
and the dark-matter halo using a Navarro-Frenk-White (NFW)
profile \citep{Navarro97}. We used the broken power law given by
\citet{Yang12} to relate the stellar mass $M_\ast$ to the halo 
mass and the mass-concentration relation given by \citet{Maccio08}
when modeling the halo.  

Both the mass-concentration and stellar-to-halo mass 
relations have significant scatter around the median 
relations. To account for this scatter, we took 10,000 
Monte Carlo realizations of lensing potentials to calculate 
the expected magnification distribution. We also drew a photo-$z$ 
and stellar mass distribution from the Monte Carlo realizations of our 
SED fits. 

Despite the proximity of the galaxy, its low mass makes magnification
a minor effect.  The median magnification from the above analysis is 
$2.8 ^{+2.3}_{-1.2}\%$,
where the lower and upper uncertainties represent the
$16^{\rm th}$ and $84^{\rm th}$ percentiles ($\pm 1\sigma$),
respectively. These models do assume a spherical NFW profile, 
but adding ellipticity to the halo does not significantly change 
our results.  This analysis shows that the systematic offset due to lensing is 
much smaller than the photometric uncertainties; we applied the lensing
correction to our derived magnitude, but it does not have a
significant effect on our distance modulus. 

\subsection{Light-curve Fit}
\label{sec:lightcurve}

\begin{figure*}
\plotone{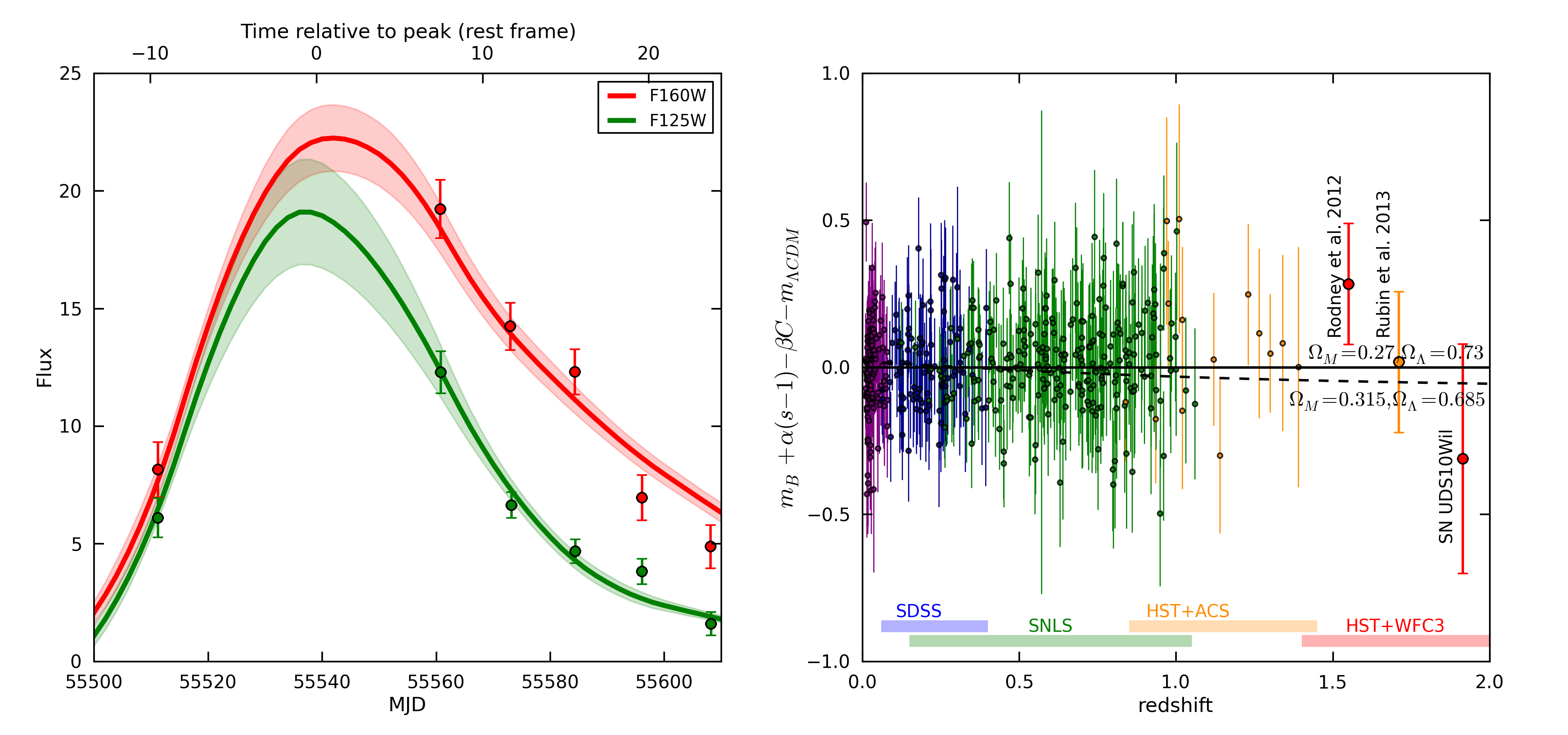}
\caption{On the left, the SALT2 light-curve fit to SN \Wilson.  SALT2 
  gives a normal set of light-curve 
  parameters along with a corrected magnitude
  of $26.15 \pm 0.39$ (consistent with $\Lambda$CDM) and a reduced
  $\chi^2$ of 1.5.  On the right, we place SN \Wilson\ on the Hubble residual
  diagram using cosmological parameters from \citet{Sullivan11} and
  H$_0 = 71.6$\,km\,s$^{-1}$\,Mpc$^{-1}$.  For comparison, we also show
  the compilation of $\sim 500$ SNe from
  \citet{Conley11}.  Lastly, using \citet{Sullivan11} values for $\alpha$ and
  $\beta$, we include the other two
  SNe\,Ia with spectroscopic evidence for classification discovered at a
  redshift greater than 1.5 \citep{Rodney12,Rubin12}.  The dotted line
  indicates the difference in $m_{\Lambda CDM}$ when using the recent cosmological parameters from the Planck
  collaboration ($\Omega_\Lambda=0.685$, $\Omega_M=0.315$; \citealp{Planck13}).}
\label{fig:hubble}
\end{figure*}

We fit the light curve using the SALT2 implementation \citep{Guy10}
contained in SNANA (Fig. \ref{fig:hubble}).  Note that although the ACS data provide a valuable color
constraint for classification purposes, we have omitted them from our
cosmological analysis.  At $z=1.914$, the ACS bands sample rest-frame
wavelengths of
2400--3300\,\AA, where SNe\,Ia are more heterogeneous \citep{Ellis08} and
may evolve with redshift \citep{Foley12}.  Furthermore, the rest-frame UV
has been problematic for SN\,Ia light-curve fitters
\citep{Kessler09b}.  Given these concerns, we discarded the ACS data
for our light-curve fitting in order to avoid introducing a bias in
the derived distance.  The ACS observations provided only a single
measurement with positive flux (F814W at MJD $=$ 55801.1), so this
does not exclude a large fraction of useful data.

The light-curve parameters for SN \Wilson\ are typical of SNe\,Ia; we
derive values of $x_1=-1.50\pm0.51$ and $C=-0.071\pm0.11$, which are
consistent with the parameters described by \citet{Kessler09b} ($\bar{C}=0.04$,
$\sigma_C=0.13$, $\bar{x}_1=-0.13$, $\sigma_{x_1}=1.24$).  SNANA
also gives a peak magnitude $m_B^*=26.20\pm0.11$.  We then converted our
SALT2 values into SiFTO values \citep{Conley08}, using the relations of \citet{Guy10}, in order to use the shape and
color constants from SNLS ($\alpha = 1.367$, $\beta = 3.179$; \citealp{Sullivan11}).  We derived a light-curve shape
and color-corrected magnitude 
($m_{\rm corr}$) using

\begin{equation}
m_{\rm corr} = m_B^* + \alpha \times (s - 1) - \beta \times C,
\end{equation}

\noindent where $m_{\rm corr}$ is equal to the
distance modulus plus the SN absolute magnitude, $M$. Here $m_B^{\ast}$ 
is the peak SN magnitude, $s$ is the SiFTO stretch parameter,
and $C$ describes the color; also, $m_B^*$ includes the lensing 
correction of $0.030 ^{+0.024}_{-0.013}$ mag.

This analysis gives a corrected magnitude of
$26.15 \pm 0.39$.  We compare to the corrected magnitude for
$\Lambda$CDM, $m_{\Lambda CDM}$,
by using  the cosmological parameters from \citet{Sullivan11}
($\Omega_\Lambda=0.73$, $\Omega_M=0.27$, $w=-1$, 
H$_0=71.6$\,km\,s$^{-1}$\,Mpc$^{-1}$) and a least-squares fit to the 
\citet{Conley11} SNe.  We added an offset of 0.27 mag to the
value of $m_B$ for the \citet{Conley11} SNe in order to match the normalization of
the SALT2 fitter contained in SNANA,
finding $m_{\Lambda CDM} = 26.46$ mag (including the offset, this gives an absolute SN magnitude
$M=-19.39$).  SN \Wilson\ is less than 1$\sigma$ from $\Lambda$CDM.

We also fit the light curve with MLCS \citep*{Jha07}, after using SDSS SNe to
determine, and correct for, the $m_{\rm corr}$ offset between MLCS and 
SALT2 fits.  MLCS gives the same corrected magnitude with a somewhat
smaller uncertainty, $m_{\rm corr}=26.15 \pm 0.27$ mag (with $\Delta = 0.30 \pm 0.18$ and 
$A_V = 0.01 \pm 0.05$\,mag).  This value is slightly brighter than expected
from $\Lambda$CDM, but consistent at 1.15$\sigma$.  We verified that the MLCS and SALT2 parameters are 
consistent with each other using relations from \citet{Kessler09b}.  

\section{Discussion}

\begin{figure}
\includegraphics[height=75mm, width=90mm]{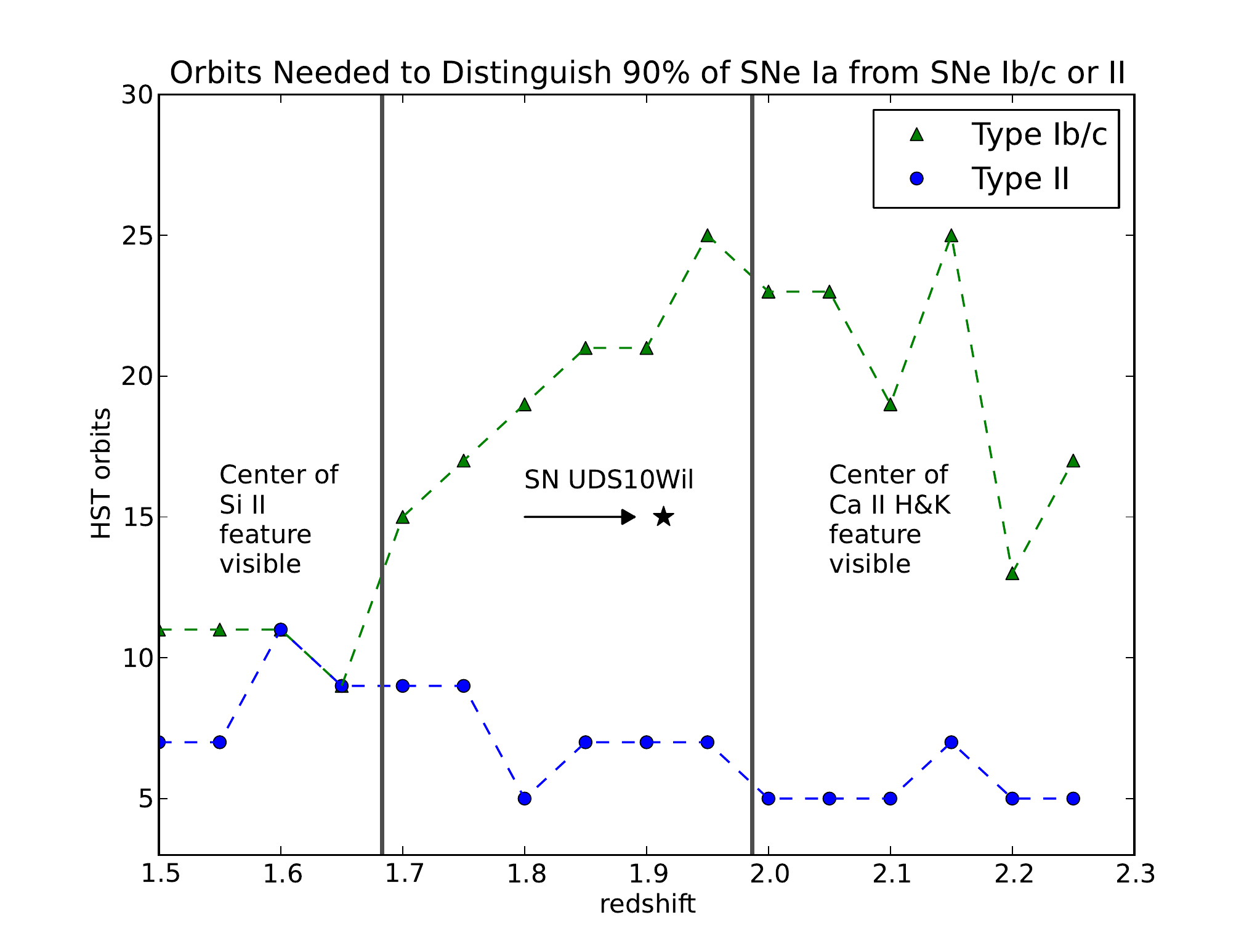}
\caption{The number of orbits necessary to rule out the possibility of
a SN\,II or SN\,Ib/c 90\% of the time when observing a SN\,Ia.  Using aXeSim,
we simulated a variety of exposure times in the redshift range 1.5--2.3.
We found that $\sim$5--10 orbits can rule out a SN\,II, but that the number
of orbits required to rule out a SN\,Ib/c possibility is significantly
greater.  The number of orbits to rule out a SN\,Ib/c is much
lower in the region where \ion{Si}{2} and \ion{Ca}{2} H\&K are completely 
visible.  These results indicate the need for photometric evidence in
SN\,Ia classification at high redshift, although the {\it HST} grism can
also be valuable in determining SN redshifts.}
\label{fig:expspec}
\end{figure}

The observations of SN \Wilson\ presented here demonstrate that the
{\it HST} WFC3 now allows the cosmological
study of SNe\,Ia
at higher redshifts than ever before.  The analysis presented above is
enabled by the photometric classification methods we employ.  However, SN 
science,
especially cosmology, has in the past relied heavily on spectroscopic
evidence for classification.

At $z > 1.5$, deriving a spectral classification with {\it HST} requires a
large number of orbits to obtain a high S/N.  In
addition, the {\it HST} IR grisms cover a relatively small rest-frame
wavelength range.  In the case of SN \Wilson, the G141 grism
wavelength range ($\sim 1.12$--1.65\,$\mu$m; rest frame 
$\sim 3840$--5660\,\AA) does not include either the \ion{Si}{2} absorption at
$\sim 6150$\,\AA\ or the \ion{Ca}{2} H\&K trough at $\sim 3750$\,\AA,
which are some of the deepest SN\,Ia features (the features have equivalent widths $\sim
100$\,\AA).  This
means that spectral classifications of high-redshift SNe using SNID
will have lower-significance correlations with SN\,Ia template spectra, and
thus often yield $r$lap values much less than the suggested minimum of
5 \citep{Blondin07}.  
 
Host-galaxy contamination can be a significant source of noise in
high-$z$ SN grism spectra.  In this work we have used aXeSim to remove
host-galaxy light from the SN+host spectrum.  However, even in a
situation where a SN is well separated from its host galaxy, spectral evidence alone may not be enough to
unequivocally classify the SN as Type Ia.  Figure \ref{fig:expspec}
shows the number of {\it HST} orbits with the G141 IR grism that are needed
for SNID to correctly distinguish a SN\,Ia from a SN\,Ib/c or a SN\,II
90\% of the time, in
the redshift range 1.5--2.3.  For this figure, we used aXeSim to
simulate 100 SN\,Ia observations per unit redshift with 2.6\,ks per
orbit, from 0 to 100 orbits (2-orbit intervals).  
The background flux was set to 95\,e$^{-}$\,s$^{-1}$, and the simulated SN
magnitude was fixed at the peak magnitude of SN Primo (an F160W Vega
magnitude of 23.98; Primo was
observed closer to maximum than \Wilson).  Each simulated grism
spectrum was then processed with SNID, and the classification was
deemed correct if the best $r$lap for a SN\,Ia template was larger than
the best SN\,Ib/c or SN\,II $r$lap (we allow $r$lap $<$ 5).

Figure \ref{fig:expspec} shows that, similar to the case of SN
\Wilson, ruling out a SN\,II possibility requires only $\sim$5--10
orbits (1.5 $< z <$ 2.3).  For a large program like CANDELS, with 200 follow-up orbits
and a desired sample of $\sim$10 SNe, this is a feasible number.
Ruling out a SN\,Ib/c, however, can require up to 25 orbits, becoming
most costly at those redshifts where
\ion{Si}{2} and \ion{Ca}{2} H\&K are not completely visible.  At very high redshift, such as $z \approx 2.2$--2.3, Figure
\ref{fig:expspec} shows that the G141 exposure time
required to distinguish a SN\,Ia from a SN\,Ib/c begins to drop.
The value of spectroscopic confirmation for such high-redshift 
SNe may warrant the necessary 
investment of orbits, especially if additional 
high-value targets can be simultaneously observed within the grism field of view.  We 
note that simulating CC\,SN observations shows that SNID can occasionally
misclassify CC\,SNe as SNe\,Ia, an effect we have not taken into
account in this analysis.

The G102 grism can also be useful for picking out features such as
\ion{Si}{2} and \ion{Ca}{2} H\&K in the
redshift ranges where the G141 grism does not contain them.  Unfortunately,
the consequence of its more limited wavelength range ($\sim
0.8$--1.15\,$\mu$m; rest frame $\sim 2750$--3950\,\AA\ at $z=1.91$)
is that a SN\,Ib/c template is more likely to match a SN\,Ia G102 spectrum well.

{\it HST} grism spectroscopy can be good at determining SN redshifts or host-galaxy 
properties.  However, for reliable SN classification, photometric
evidence is often important.  In the case of SN \Wilson\, we improved upon the
photometric methods of \citet{Rodney12} by introducing a quantitative
Bayesian method that returns probabilities for each SN type.  With
both photometric and spectroscopic methods, we can be confident in our
classification and subsequent analysis.

\section{Conclusions}

At a redshift of 1.914, SN \Wilson\ is the most distant SN\,Ia yet
known.  Classification of this SN rests on photometry and grism
spectroscopy, which rules out the possibility of a CC SN.  
The spectral evidence alone disfavors the possibility of a SN\,II, while
supporting a SN\,Ia or SN\,Ib/c hypothesis.  The combined SN
colors and rapid decline rate are inconsistent with a
CC SN and in good agreement with a SN\,Ia model.  

We find that SN \Wilson\ is not significantly lensed, and its light-curve 
fit (with SALT2) is consistent with $\Lambda$CDM.  An alternative fit with MLCS \citep{Jha07} 
is slightly brighter than $\Lambda$CDM, but consistent at
1.15$\sigma$.  


When the full analysis of the CANDELS
SNe is complete and combined with the data from the Cluster Lensing
and Supernova survey with Hubble (PI: Postman; \citealp{Postman12}), 
we expect that SN \Wilson\ will be one of a sample of $\sim 10$ SNe\,Ia 
above a redshift of 1.5 to be 
found by these programs.  This SN is an example of an object in a new area of 
SN cosmology, one which has only begun to be explored in the 
last few years with the advent of WFC3 on {\it HST} and one with unique
classification challenges.  However, with the full sample of SNe at 
redshift greater than 1.5, new limits on 
the evolution of dark energy, the DTD, and the evolution 
of the SN\,Ia population will become possible.

\acknowledgements

We thank the anonymous referee for many helpful
comments, and Masao Sako and Rick Kessler for their invaluable
assistance with SNANA and PSNID.  In addition, our aXe and aXeSim analysis was made possible by help
from Jeremy Walsh, Harold Kuntschner,
Martin Kummel, Howard Bushouse, and Nor Pirzkal.  We 
also thank Daniel Scolnic for many useful discussions.  
Financial support for this work was provided by NASA through grants
HST-GO-12060 and HST-GO-12099 from the Space Telescope Science
Institute, which is operated by Associated Universities for Research
in Astronomy, Inc., under NASA contract NAS 5-26555.
Additional support for S.R. was provided by NASA through Hubble Fellowship 
grant HST-HF-51312. A.V.F. is also grateful for the support 
of National Science Foundation (NSF) grant AST-1211916, the
TABASGO Foundation, and the Christopher R. Redlich Fund.  
 The Dark Cosmology Centre is funded by the DNRF.  R.P.K.
was supported in part by NSF 
grant PHY11-25915 to the Kavli Institute for Theoretical 
Physics at the University of California, Santa Barbara.  
O.G. acknowledges support by a grant from the Israeli Science Foundation.

{\it Facilities:} \facility{HST (WFC3, ACS)} \facility{VLT:Kueyen (X-shooter)}

\appendix
\section{Photometric Classification Method}

We began our classification procedure by using SNANA \citep{Kessler09}
to generate a Monte Carlo simulation of
30,000 SNe at redshift 1.91.  10,000 simulated SNe\,Ia
were based on the SALT2 model \citep{Guy10}, with values of
the shape parameter $x_1$ drawn uniformly in the range $-3$ to 3 and the
color parameter $C$ from $-0.4$ to 0.6. These ranges cover the observed
distribution of SALT2 parameter values \citep{Kessler09b}, and the
$C$ term accounts for both intrinsic SN color and host-galaxy extinction
\citep{Guy07}.  The remaining 20,000 simulated SNe were split
evenly between the two principal \CCSN\ classes, with light curves
based on 16 Type Ib/c and 27 Type II SN templates that comprise the
SNANA non-SN\,Ia library (including subtypes Ib, Ic, II-P, IIn, and
II-L).  Host-galaxy reddening was applied to each
simulated \CCSN\ using \Rv\ = 3.1, with a random draw of \Av\ in the
range 0 to 7\,mag using the \citet{Cardelli89} reddening law.  
For both the \SNIa\ and the \CCSN\ 
simulations we chose random values for the date of the light-curve 
peak, using a range spanning the first to the second epoch of
\Wilson\ observations.

To compare each of the 30,000 synthetic light curves to the $N=15$
photometric observations of SN \Wilson, we computed the $\chi^2$
statistic given by

\begin{equation}
\chi^2 = \sum\limits_{i=1}^{N}\frac{(F_{{\rm obs},i} - A\times F_{{\rm sim},i})^2}{\sigma_{{\rm obs},i}^2+\sigma_{{\rm sim},i}^2},
\end{equation}

\noindent where $F_{{\rm obs},i}$ and $\sigma_{{\rm obs},i}$ are the fluxes 
and uncertainties for each observation. Here $F_{{\rm sim},i}$ and 
$\sigma_{{\rm sim},i}$ are the 
fluxes and uncertainties (respectively) for a single simulated SN on each
observation date.  The variable $A$ is a scaling parameter, described below.  For SNe\,Ia, most of their intrinsic variability 
can be described by the SALT2 model's shape and color parameters.
Additional variability causes scatter about
the Hubble diagram, and is 
given by \citet{Guy10} as 8.7\% in distance modulus.  
We treat this variability as approximately equal to the model uncertainty, 
which in flux space translates to
$\sigma_{{\rm simIa},i} = 0.08A \times F_{{\rm sim},i}$.  

CC\,SNe have greater heterogeneity,
such that our relatively small set of discrete
templates cannot describe the entire population.  By setting a
nonzero $\sigma_{{\rm simCC},i}$, our limited CC\,SN template library can 
more accurately represent this diverse class.  Considering a similar 
problem, \citet{Rodney09} estimated $\sigma_{{\rm simCC},i}$ by
measuring the flux difference
between all possible pairwise comparisons of templates of the same
subclass and taking the median.  
We classify SN \Wilson\ using more
templates than \citet{Rodney09}, such that the CC\,SN population is better
sampled and less uncertainty is present.  However, we adopt their
value of $\sigma_{{\rm simCC},i}=0.15A \times
F_{{\rm sim},i}$ as a conservative estimate.

We next chose the optimal distance or absolute
magnitude of every simulated SN, therefore removing the assumptions
on cosmology and SN luminosity functions that are built into the SNANA
simulations.  Here we have multiplied $F_{{\rm sim},i}$ by $A$, a free
parameter that introduces a coherent flux scaling across all bands.
We find a separate value for $A$ with each of the 30,000 comparisons,
using $\chi^2$ minimization to match the simulated magnitudes to the
data with the equation

\begin{equation}
A=\frac{\sum\limits_{i=1}^NF_{{\rm sim},i}F_{{\rm obs},i}/\sigma_{{\rm obs},i}^2}{\sum\limits_{i=1}^NF_{{\rm sim},i}^2/\sigma_{{\rm obs},i}^2}.
\end{equation}

We then converted the measured $\chi^2$ values into a Type Ia SN
classification probability using a simple Bayesian framework, similar
to \citet{Poznanski07}, \citet{Kuznetsova07}, and \citet{Sako11}.  
The {\em likelihood} that
the data ($D$) match a simulated SN of type $T_j$ with
parameters $\bm{\theta}$ (shape $x_1$, color $C$ for SN Ia or $A_V$ for
CC\,SN types, and time of
maximum light) is given by

\begin{equation}
p(D|\bm{\theta},T_j) = \frac{e^{-\chi^2/2}}{\prod\limits_{i=1}^{N}\sqrt{2\pi (\sigma_{{\rm obs},i}^2+\sigma_{{\rm sim},i}^2)}} , 
\end{equation}

\noindent where $\chi^2$ is given in Equation 1.  Multiplying by prior 
probability distributions for each of the model parameters then gives us 
the {\em posterior probability} for each point in parameter space,
$p(\bm{\theta}|T_j)p(D|\bm{\theta},T_j)$.  As we are interested in model
selection, not parameter estimation, we can marginalize over all of the
nuisance parameters $\bm{\theta}$.  Approximating the marginalization
integral with a discrete sum, the probability of SN type $T_j$
given the model is

\begin{equation}
p(D|T_j)=\sum\limits_{i=1}^{N_{\rm sim}(T_j)}p(\bm{\theta}|T_j)p(D|\bm{\theta},T_j)\delta\bm{\theta}.
\end{equation}

\noindent For SN\,Ia parameters $x_1$ and $C$, we applied Gaussian
priors based on the values given by \citet{Kessler09b} ($\bar{C}=0.04$, 
$\sigma_C=0.13$, $\bar{x}_1=-0.13$, $\sigma_{x_1}=1.24$).  For the CC SNe
parameter $A_V$,
we used the Monte Carlo recipe provided by \citet{Riello05} and implemented by 
\citet{Dahlen12} for a random galaxy orientation.  The distribution is
peaked at $A_V=0$\, mag and falls off quickly such that $A_V \gtrsim 3$\,mag 
is very unlikely.  We used a flat prior for the time of peak
brightness.

Note that for computational efficiency we have used SNANA to
sample the multi-dimensional model parameter space using a Monte Carlo
simulation with uniform sampling distributions (instead of the more
typical approach, using a grid of parameter values). Thus, we must
approximate $\delta\bm{\theta}$ -- the vector of step sizes along each
dimension of parameter space -- using the range over which each
parameter is sampled: 

\begin{equation}
\delta\bm{\theta}=\frac{1}{N_j}\prod\limits_{k=1}^{N_\theta}\Delta\theta_k,
\end{equation}

\noindent where $\Delta\theta_k$ is each range, $N_j$ is the
number of simulated SNe in the class (we used 10,000), and the product
is over $N_\theta$,
the number of parameters $\theta$ for the model: 3 for SNe\,Ia
($x_1$, $c$, $t_{pk}$) and 2 for CC\,SNe ($A_V$, $t_{pk}$). 

Lastly, we multiplied each model by a SN rate prior $P(T_j)$.  This
prior is the fraction of SNe at redshift 1.91 that we expect to be a
given type.  We began by adopting the rate measurement from
\citet{Dahlen08} for SNe\,Ia and the local rates from \citet{Li11}
for CC\,SNe.  We scaled the CC\,SN rates according to the cosmic 
star-formation history of \citet{Hopkins06} using the form of
\citet{Cole01} and a modified Salpeter IMF
\citep{Baldry03}.  The normalized rates showed, as an estimate, that
one could expect only $\sim 2$\% of SNe at this redshift to be of Type
Ia.  The SN \Wilson\ host galaxy SED (\S3.2) is consistent with a
starburst galaxy, so it is possible that these average rates
overestimate the SN\,Ia rate in this galaxy.  In addition, SN rates are very uncertain at this redshift, and the \citet{Dahlen08}
rates at this redshift are consistent with 0 SNe Ia; however, 
we note that lowering this rates prior by an order of magnitude
still returns a classification
probability greater than 99\%.  Thus the result is largely independent of the
uncertainty in SN\,Ia rates.

Applying Bayes' theorem gives the final probability that SN \Wilson\
is of Type Ia:

\begin{equation}
p(Ia|D) = \frac{p(D|Ia)p(Ia)}{\sum\limits_{j}p(D|T_j)p(T_j)},
\end{equation}

\noindent where the summation is over Type Ia, Ib/c, and II SN
models ($T_j$).  In the case of SN \Wilson, our likelihood function is sufficiently narrow that the priors
have only a minor effect.  Thus, we found that allowing $R_V$, SALT2 
parameters $\alpha$ and $\beta$, or parameter ranges to vary does not 
substantially alter the high probability that this SN is of Type Ia.

\bibliographystyle{apj}

\end{document}